\documentclass[fleqn,usenatbib]{mnras}

\usepackage{graphicx}	
\usepackage{amsmath}	
\usepackage{amssymb}	

\usepackage{csquotes}
\usepackage{bm}
\usepackage{subcaption}
\usepackage{booktabs}
\usepackage{grffile}
\usepackage{overpic}
\usepackage{float}
\usepackage{enumitem}
\usepackage{amsfonts}
\usepackage{easy-todo}
\usepackage{listings}
\usepackage{lineno}
\usepackage[normalem]{ulem}

\hypersetup{
    colorlinks=true,
    linkcolor=blue,
    filecolor=magenta,      
    urlcolor=cyan,
}
 
\newdimen\figrasterwd
\figrasterwd\textwidth

\newcommand{\rev}[1]{{{#1}}}
\newcommand{\newold}[2]{{#1}{}}
\newcommand{\revcite}[1]{{#1}}

\newcommand{\ten}[1]{{\bm #1}}
\renewcommand{\vec}[1]{{\bm #1}}

\newcommand{\mycaption}[2]{\caption[#1]{\emph{#1} #2}}

\title{Flute and kink instabilities in a dynamically twisted
    flux tube with anisotropic plasma viscosity}

\author[J.~Quinn and R.~Simitev]{
James J.~Quinn,$^{1}$\thanks{E-mail: jamiejquinn@jamiejquinn.com}
Radostin D.~Simitev$^{2}$
\\
$^{1}$ Research Software Development Group, University College London, Gower Street, London WC1E 6BT, UK \\
\href{https://orcid.org/0000-0002-0268-7032}{orcid.org/0000-0002-0268-7032} \\
$^2$ School of Mathematics and Statistics, University of Glasgow,
Glasgow G12 8QQ,
UK \\ \href{https://orcid.org/0000-0002-2207-5789}{orcid.org/0000-0002-2207-5789}\\
}

\date{Accepted XXX. Received YYY; in original form ZZZ}

\pubyear{2021}

\begin{document}
\label{firstpage}
\pagerange{\pageref{firstpage}--\pageref{lastpage}}
\maketitle

\lstset{
  language=[90]Fortran,
  basicstyle=\ttfamily,
  keywordstyle=\color{red},
  commentstyle=\color{green},
  frame=single
}

\graphicspath{{images/kink_instability_straight/}}

\begin{abstract}
Magnetic flux tubes such as those in the solar corona are subject to a number
of instabilities. Important among them is the kink instability which plays
a central part in the nanoflare theory of coronal heating, and for this reason
in numerical simulations it is usually induced by tightly-controlled
perturbations and studied in isolation. In contrast, we find that
fluting modes of instability are readily excited when disturbances are
introduced in our magnetohydrodynamic flux tube simulations by dynamic
twisting of the flow at the boundaries. We also find that the flute instability, which has been 
theorised but rarely observed in the coronal context,  is strongly enhanced
when plasma viscosity is assumed anisotropic. We proceed to investigate the
co-existence and competition between flute and kink instabilities for a range
of values of the resistivity and of the parameters of the anisotropic and
isotropic models of viscosity. We conclude that while the flute instability
cannot prevent the kink from ultimately dominating, it can significantly delay
its development especially at strong viscous anisotropy induced by intense
magnetic fields.
\end{abstract}

\begin{keywords}
Sun: corona -- Sun: magnetic fields -- instabilities -- plasmas -- MHD
\end{keywords}

\section{Introduction}




The helical kink instability is a form of ideal magnetohydrodynamic (MHD)
instability which occurs in highly twisted magnetic flux tubes such as those
making up much of the solar corona~\citep{realeCoronalLoopsObservations2014}
and has been well studied in the coronal
context
\citep{hoodKinkInstabilitySolar1979,
  hoodCoronalHeatingMagnetic2009,
  browningSolarCoronalHeating2003b,
  Torok2003, Torok2004, Torok2005,
  barefordShockHeatingNumerical2015,
  quinnEffectAnisotropicViscosity2020}.
Given its energetic nonlinear development, it is considered a potential
mechanism for heating the solar corona through the theory of
nanoflares~\citep{klimchukSolvingCoronalHeating2006,browningMechanismsSolarCoronal1991}
and a key mechanism in the production of solar
flares~\citep{hoodKinkInstabilitySolar1979}. \rev{Our previous work}
investigated a twisted magnetic flux tube already linearly unstable to
the 
helical kink instability, focussing specifically on the effect of anisotropic
viscosity on the nonlinear
dynamics~\citep{quinnEffectAnisotropicViscosity2020}.
\rev{There and in} most other investigations of
the kink instability e.g.~that of \revcite{\citet{hoodCoronalHeatingMagnetic2009}},
a perturbation is applied to an already significantly twisted flux tube. An
alternative way to excite the instability (and the way employed here) is to
start with an initially straight field and apply twisting motions at the
boundaries to form a twisted flux tube which eventually becomes unstable. This
kind of dynamic excitation of the kink instability \rev{}
represents more closely the actual evolution of magnetic flux tubes and the
associated instabilities in the solar corona. In our simulations, the dynamic
twisting of the flux tube reveals an additional instability, the flute
instability, which has been theorised, for example
by~\revcite{\citet{priestMagnetohydrodynamicsSun2013}}. While oscillations resembling
flute perturbations have been found in simulations of coronal
loops~\citep{terradasEffectMagneticTwist2018}, to our knowledge, this is the
first time the flute instability has been investigated computationally in
a coronal context.

The flute instability arises in magnetised plasmas where the plasma pressure
gradient is oriented in the same direction as the field line curvature, that is
the pressure and magnetic tension forces compete. This is similar to the
competition between pressure and gravitational forces which gives rise to the
Rayleigh-Taylor instability (RTI). In magnetohydrodynamic terminology, the RTI
is a typical example of an ideal interchange instability, where magnetic field
lines are minimally bent and are, instead, exchanged during the evolution of
the instability. The ideal flute instability is another example of an ideal
interchange instability but confined to a cylindrical geometry, the term
``flute instability'' referring to its likeness to a fluted column. In
a twisted flux tube like a simple, unbraided coronal loop, the magnetic
curvature is always directed towards the axis so the tube may be unstable to
fluting when the pressure decreases outwards from a high-pressure core. Such
a pressure distribution is generated in the flux tubes studied here as a result
of the driving. The appearance of the flute instability is illustrated by, for
example, the pressure contours in Figure~\ref{fig:kink_field_line_plots}, where
the perturbations follow the pitch of the twisted field. 

In other solar contexts, interchange instabilities can be found in the form of
ballooning modes in arcades~\citep{hoodBallooningInstabilitiesSolar1986}, as
the instability which forms tubes of specific size in the
photosphere~\citep{bunteInterchangeInstabilitySolar1993}, and in the buoyancy of
flux tubes~\citep{schuesslerInterchangeInstabilitySmall1984}. However, the
flute instability specifically is more commonly studied in fusion
contexts~\citep{mikhailovskiiInstabilitiesConfinedPlasma1998,zhengAdvancedTokamakStability2015,wessonHydromagneticStabilityTokamaks1978}.
In fusion, the focus is generally on understanding how a particular plasma
device may be stabilised to the instability in particular geometries such as
that of the mirror machine~\citep{jungwirthTheoryFluteInstability1965} or in
toroidal geometries such as the
tokamak~\citep{shafranovFluteInstabilityCurrentcarrying1968}. The resistive
flute instability (also known as the resistive interchange instability) can be
excited even when the ideal flute instability is stabilised. As a result, this
has been given significantly more
attention~\revcite{\citep{johnsonResistiveInterchangesNegativeV1967,correa-restrepoResistiveBallooningModes1983}}.
While this body of research is useful and applicable in solar contexts, it is
mostly limited to the study of the stability and linear development of the
flute instability, the nonlinear development being of secondary importance in
the investigation of fusion devices. More detailed investigations of its
nonlinear development is required to understand its importance in the context
of coronal dynamics and coronal heating. The development of the flute
instability and its interaction with the simultaneously growing kink
instability is the main focus of this work and the experiments described here
represent an initial exploration into the nonlinear flute instability in the
solar corona.

In addition to our main goal, of particular interest here is
the effect of
anisotropic plasma viscosity, which in the following is found to strongly
influence the growth of the flute instability. It is well known that viscosity
in magnetised plasmas (such as those which make up the solar corona) is
anisotropic and strongly dependent on the strength and direction of the local
magnetic
field~\citep{hollwegViscosityChewGoldbergerLowEquations1986,hollwegViscosityMagnetizedPlasma1985,braginskiiTransportProcessesPlasma1965}.
To take this into account,
\revcite{\citet{mactaggartBraginskiiMagnetohydrodynamicsArbitrary2017}}
developed a phenomenological model of anisotropic  viscosity
that captures the main physics of viscosity in the solar corona as outlined in the
analysis of~\cite{braginskiiTransportProcessesPlasma1965}, namely parallel
viscosity in regions of strong field  strength and isotropic viscosity in
regions of very weak or zero field strength. For brevity, we will refer to this
model of viscosity as ``the switching model''.
In \revcite{\citet{quinnEffectAnisotropicViscosity2020,quinnKelvinHelmholtzInstabilityCollapse2021}}, 
we implemented the switching model as a module for the widely-used general MHD
code Lare3d~\citep{arberStaggeredGridLagrangian2001}, and demonstrated
significant effects of anisotropic viscosity on the development of the
nonlinear MHD kink instability and the Kelvin-Helmholtz instability. More
generally, the interest in anisotropic viscosity stems from the open question
of which heating mechanism (viscous or Ohmic) is dominant in the solar
corona~\citep{klimchukSolvingCoronalHeating2006}, an important facet of solving
the coronal heating problem. Using scaling laws, it has been suggested that
viscous heating (generated through anisotropic viscosity) can dwarf that of
Ohmic
heating~\citep{craigAnisotropicViscousDissipation2009a,litvinenkoViscousEnergyDissipation2005}.
However, due to computational and observational limitations, this cannot be
directly tested, and so the influence of other factors such as small scale
instabilities and turbulence is relatively
unknown~\revcite{\citep{klimchukSolvingCoronalHeating2006}}. In addition to directly
heating the plasma, viscosity plays a part in the damping of instabilities and
waves~\citep{rudermanSlowSurfaceWave2000}. It is this effect we are most
interested in here, and it shall be reported that the use of anisotropic
viscosity permits the \rev{growth} of the flute instability, which is otherwise
strongly damped by isotropic viscosity.

\rev{The value of plasma resistivity also affects the
development of the flute instability because pressure gradients
generated through Ohmic heating substantially contribute to its growth.
Ideally, our simulations would be performed using a realistic coronal
resistivity values of approximately
$10^{-8}$~\citep{craigAnisotropicViscousDissipation2009a}, however, due
to the dissipative nature of numerical schemes (particularly when using shock
capturing techniques) this is computationally infeasible.
To overcome this limitation, we perform and compare simulations at two computationally
accessible resistivity values, $10^{-3}$ and $10^{-4}$,
in an attempt to extrapolate results towards more realistic values.
This comparison runs as an additional theme of the paper,
if not a primary aim.}
%

Our article is organised as follows. Section~\ref{sec-flute-intro} introduces
the flute instability and \rev{recalls relevant background on its linear stability analysis}.
Section~\ref{sec-numerical-setup} describes the governing equations,
the coronal loop model and the numerical parameters used. Section~\ref{sec-results}
presents the overall development of the flute instability before
\rev{proceeding to compare simulations in the cases of various viscous
anisotropy and various resistivity values. Here results are
organised by resistivity values as this allows to contrast isotropic
and anisotropic viscosity cases more directly.}
Section~\ref{sec-discussion} discusses the limitations of the simulations, with
suggestions for future work, and section~\ref{sec-conclusions} presents our
conclusions in the wider context of coronal heating.

\section{The flute instability}
\label{sec-flute-intro}

In general, the stability of a cylindrical twisted magnetic flux tube is analysed using perturbations of the form
\begin{equation}
  \label{eq:kink_perturbation}
\xi(r, \theta, z) = \xi(r) e^{i(m\theta + kz -\omega t)},
\end{equation}
where $\omega$ is the oscillation frequency in time $t$, $m$ and $k$
are the wavenumbers in the azimuthal and axial directions, $\theta$
and $z$, respectively and $r$ is the radial coordinate in cylindrical 
polars. The helical kink instability occurs for perturbations where $m=1,
k\ne0$ and is the only instability of this form which is a body instability,
i.e.{} it moves the entire body of the flux tube. Perturbations where $m>1$
are termed flute or interchange instabilities.

When the magnetic field is sheared, as in a twisted magnetic flux tube, an
interchange instability (such as the flute instability) is confined to
a surface where the peaks and troughs follow the shear of the field. That is,
the instability is confined to the surface where the perturbation wavevector
$(0, m/r, k)$ is perpendicular to the direction of the field, known as the
``resonance surface''. In an axisymmetric twisted flux tube the resonance
surface is located at a radius $r$ specified by
\begin{equation}
  \label{eq:resonant_surface}
\frac{m}{r} B_{\theta}(r) + kB_z(r) \approx 0.
\end{equation}

The stability of an infinite cylindrical flux tube to perturbations of the
form~\eqref{eq:kink_perturbation} is given by the classical Suydam's
criterion~\citep{suydamStabilityLinearPinch1958}
\begin{equation}
  \label{eq:suydams_criterion}
\frac{B_z^2 S^2}{4} + 2 r p' > 0,
\end{equation}
where $S = r q'/q$ is a measure of the shear, $q = 2\pi r B_z / L B_{\theta}$
is the safety factor for a flux tube of length $L$ and a prime denotes
differentiation with respect to
$r$~\citep{mikhailovskiiInstabilitiesConfinedPlasma1998}. This applies to both
flute and kink instabilities although many additional effects such as
line-tying are not incorporated into the corresponding linear analysis. The
effect of line-tying on the kink instability is investigated
by \revcite{\citet{hoodKinkInstabilitySolar1979}}. Where~\eqref{eq:suydams_criterion} is
not satisfied, the flux tube may be unstable to perturbations of the
form~\eqref{eq:kink_perturbation}. When $m>1$, the perturbations remain local
to resonant surfaces given by~\eqref{eq:resonant_surface}. When Suydam's
criterion is satisfied and the flux tube is linearly stable, it may still be
unstable to non-local perturbations, where the shear and pressure are small
enough that interchange perturbations do not need to
satisfy~\eqref{eq:resonant_surface}. Additionally, the inclusion of resistivity
generally reduces the stabilising effect of the shear, permitting growth of
a resistive interchange mode, albeit at a significantly slower rate than that of the ideal
instability
\citep{mikhailovskiiInstabilitiesConfinedPlasma1998}. \rev{At the
  values of resistivity studied here, the resistive growth rate is
  expected to be approximately two orders of magnitude less than the corresponding ideal rate.} Furthermore, it will be
found that the ideal linear analysis
of~\cite{mikhailovskiiInstabilitiesConfinedPlasma1998} is sufficient for
understanding the flute instabilities investigated here since the associated
flux tubes substantially fail the criterion~\eqref{eq:suydams_criterion}. \rev{For these reasons, we consider only the ideal flute instability.}

While Suydam's condition gives an indication of the stability of a flux tube to
a given perturbation, the linear growth rate of the ideal flute instability
$\gamma$, \rev{defined as the imaginary part of $\omega$ in ansatz
  \eqref{eq:kink_perturbation},} can be \rev{determined} via a stability analysis
  analogous to that of the Rayleigh-Taylor
  instability~\citep[][]{goldstonIntroductionPlasmaPhysics2020}.
 \rev{The fastest growing mode in the $r-$direction is found to be the longest
wavelength mode, while the fastest growing modes in the $\theta-$direction
are found to be those with the shortest wavelengths, i.e.~large values of $m$ in
the notation of equation \eqref{eq:kink_perturbation}. In particular, for
all modes with wavelengths in the $\theta-$direction that are shorter
than both the pressure-gradient scale-length and the radial height
of the plasma, the growth rate $\lambda$ tends to the limit}  
\begin{equation}
  \label{eq:fluting_growth_rate}
\gamma^2 = \frac{2|\nabla p|}{\rho R_c},
\end{equation}
where $R_c$ is the radius of curvature of the magnetic field.
\rev{We find that this expression gives a good
  estimate of the growth rate of the flute instability in our
  numerical simulations even at moderate values of $m$.} 
Equation \eqref{eq:fluting_growth_rate}
only applies when the pressure gradient and radius of curvature vector are in
the same direction, that is the plasma is constrained by a concave magnetic
field such that the pressure forces and magnetic tension forces are in
competition. In a cylindrical, twisted flux tube, the field is always concave
towards the central axis of the tube, so any inwardly directed pressure
gradient is potentially unstable to fluting.

Throughout this paper, the twisted flux tube generated by the drivers has
a pressure profile which is approximately axisymmetric, and independent of $z$
away from the boundaries at $z=\pm2$, and has a negative gradient, hence
$|\nabla p|$ may be written as $-d p/ dr$. Similarly, away from the boundaries,
the magnetic field has a negligible $r$ component and little dependence on
$\theta$ and $z$, allowing the field to be approximated as $\vec{B} = (0,
B_{\theta}(r), B_z(r))^{\text{T}}$, in cylindrical coordinates $(r, \theta,
z)$. For a twisted field of this form, the radius of curvature is given by
\begin{equation} \label{eq:radius_of_curvature} R_c
= \frac{1}{|(\vec{b}\cdot\nabla) \vec{b}|} = \frac{r}{b_{\theta}^2},
\end{equation} where $\vec{b} = \vec{B}/|\vec{B}|$ is the unit vector in the
direction of the magnetic field and $b_{\theta}$ is the component of $\vec{b}$
in the azimuthal direction. \rev{These approximations allow} the growth rate to be
written as
\begin{equation}
\label{eq:fluting_growth_rate2}
\gamma_{\text{ideal}}^2 = \frac{-2p'}{\rho R_c}.
\end{equation}
This approximation for the growth rate continues to hold while the flux tube
remains relatively axisymmetric, that is while the kink instability remains in
its linear phase.

The stability criterion \eqref{eq:suydams_criterion} and the linear growth rate
approximation \eqref{eq:fluting_growth_rate2} are useful only as a guide and
for approximate analysis of the numerical simulations presented in this work.
The precise form of the equilibrium state and the perturbations needed for the
validity of \eqref{eq:suydams_criterion} and \eqref{eq:fluting_growth_rate2}
were used by~\revcite{\citet{quinnEffectAnisotropicViscosity2020}}. In contrast, in the
experiments reported in the following the system is driven and instabilities
occur spontaneously due to random perturbations. As a result of the driving, the
flux tube is also not in static equilibrium initially.

\section{Mathematical formulation and numerical setup}
\label{sec-numerical-setup}

We consider the magnetohydrodynamic equations for the density $\rho$, plasma
velocity $\vec{u}$, pressure $p$, magnetic field $\vec{B}$ and internal energy
$\varepsilon$, in their non-dimensionalised visco-resistive form
\begin{subequations}
  \label{eq:MHD}
  \begin{gather}
\label{eq:mhda}
\frac{D\rho}{Dt} = - \rho \vec{\nabla} \cdot \vec{u},\\
\rho\frac{D\vec{u}}{Dt} = -\vec{\nabla} p + \vec{\jmath} \times \vec{B} + \vec{\nabla} \cdot \ten{\sigma},\\
\frac{D\vec{B}}{Dt} = (\vec{B} \cdot \vec{\nabla})\vec{u} - (\vec{\nabla} \cdot \vec{u})\vec{B} + \eta \nabla^2 \vec{B},\\
\rho\frac{D\varepsilon}{Dt} = -p \vec{\nabla} \cdot \vec{u} + {Q}_{\nu} + {Q}_{\eta},
    \end{gather}
\end{subequations}
where $\eta$ is the non-dimensionalised resistivity, $\vec{\jmath} = \nabla
\times \vec{B}$ is the current density and the terms ${Q}_{\nu} = \ten{\sigma}
: \vec{\nabla}\vec{u}$ and ${Q}_{\eta} = \eta | \vec{\jmath} |^2$ are viscous
heating and Ohmic heating, respectively. The system is closed by the inclusion
of the equation of state for an ideal gas 
\begin{equation}
\varepsilon = \frac{p}{\rho(\gamma - 1)},
\end{equation}
with the specific heat ratio is given by $\gamma = 5/3$.

Two different models for the viscosity stress tensor $\ten{\sigma}$ will be
compared and contrasted in this study. The first model is the conventional
isotropic (or Newtonian) viscosity stress tensor used in the vast majority of
the existing literature, so that, 
\begin{equation}
  \label{eq:isotropic_viscous_tensor}
\ten{\sigma = }\ten{\sigma}_{\text{iso}} = \nu \ten{W},
\end{equation}
where $\nu$ is the viscous transport parameter, generally referred to as the
viscosity,
\begin{equation}
  \label{eq:rate_of_strain}
  \ten{W} = \nabla\vec{u} + (\nabla\vec{u})^T - \tfrac{2}{3}(\nabla \cdot \vec{u})\ten{I},
\end{equation}
is the rate of strain tensor, and $\ten{I}$ is the  $3\times 3$ identity
tensor. The second model, which is the one of actual interest, is the
anisotropic viscosity stress tensor given by
\begin{equation}
  \label{eq:switching_model}
\ten{\sigma} = \ten{\sigma}_\text{aniso} = \nu \left[\frac{3}{2}(\ten{W}\vec{b}\cdot\vec{b}) \left( \vec{b} \otimes \vec{b} - \frac{1}{3}\ten{I} \right)\right],
\end{equation}
where $\vec{b}$ is the unit vector in the direction of the magnetic
field.

Expression \eqref{eq:switching_model} is identical to the strong field
approximation of the general anisotropic viscosity tensor derived
by \revcite{\citet{braginskiiTransportProcessesPlasma1965}}. Expressions
\eqref{eq:isotropic_viscous_tensor} and \eqref{eq:switching_model} arise as
asymptotic limits of the more general switching model used in our earlier works~\citep{mactaggartBraginskiiMagnetohydrodynamicsArbitrary2017,quinnEffectAnisotropicViscosity2020,quinnKelvinHelmholtzInstabilityCollapse2021}
which includes both isotropic and anisotropic contributions and can switch
gradually between them depending on the strength of the magnetic field at
a given spacio-temporal location. For example, in the vicinity of a null point
where the magnetic field becomes weak the isotropic viscosity contribution
becomes dominant in the switching model. Switching between the two limit cases
is not relevant in the present study where the variations in the magnetic field
are not significantly large.

The non-dimensionalisation of equations \eqref{eq:MHD} is identical to that
used in the earlier
works by \revcite{\citet{quinnEffectAnisotropicViscosity2020,quinnKelvinHelmholtzInstabilityCollapse2021}}
.
A typical magnetic field strength $B_0$, density $\rho_0$ and length scale
$L_0$ are chosen and the other variables non-dimensionalised appropriately.
Velocity and time are non-dimensionalised using the Alfv\'en speed $u_A = B_0
/ \sqrt{\rho_0 \mu_0}$ and Alfv\'en crossing time $t_A = L_0/u_A$,
respectively. Temperature is non-dimensionalised via $T_0 = u_A^2 \bar{m}
/ k_B$, where $k_B$ is the Boltzmann constant and $\bar{m}$ is the average mass
of ions, here taken to be $\bar{m} = 1.2m_p$ (a mass typical for the solar
corona) where $m_p$ is the proton mass. Dimensional quantities can be recovered
by multiplying the non-dimensional variables by their respective reference
value (e.g. $\vec{B}_{\dim} = B_0 \vec{B}$). The reference values used here are
$B_0 = 5 \times 10^{-3}$ T, $L_0 = 1$ Mm and $\rho_0 = 1.67 \times 10^{-12}
\ \text{kg m}^{-3}$, giving reference values for the Alfv\'en speed $u_A
= 3.45\ \text{Mm s}^{-1}$, Alfv\'en time $t_A =
0.29\ \text{s}$ and temperature $T_0 = 1.73 \times 10^{9}\ K$.      

The following initial and boundary conditions are used to form a magnetic flux
tube and excite instabilities by dynamic twisting. The magnetic field is
prescribed as initially straight and uniform,
\begin{equation}
\vec{B} = (0, 0, 1)^{\text{T}},
\end{equation}
in a cube of size $[-2,2]^3$, with further test simulations run using an
elongated domain of size $4\times4\times20$. Initially, the velocity is set
everywhere to $\vec{u} = \vec{0}$, the density to $\rho = 1$, and the internal
energy to $\varepsilon = 8.67\times 10^{-4}$. This corresponds to a typical
coronal temperature of $10^6$ K and a plasma beta of $1.11 \times 10^{-4}$. At
the boundaries, the magnetic field, velocity, density, and energy are fixed to
their initial values and their derivatives normal to the boundaries are set to
zero  except where twisting velocity ``driver'', described below, is
prescribed.

\begin{figure}
  \centering
  \begin{subfigure}{.49\textwidth}
  \centering
  \includegraphics[width=1.0\linewidth]{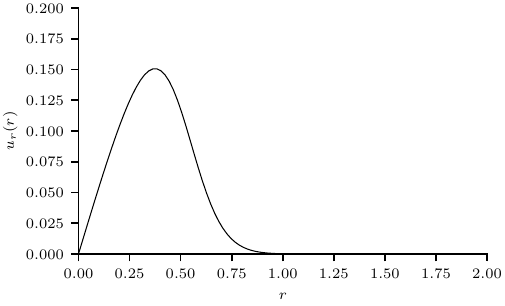}
  \caption{Radial dependence of driver}
  \label{fig:kink_radial_driver}
  \end{subfigure}
  \begin{subfigure}{.49\textwidth}
  \centering
  \includegraphics[width=1.0\linewidth]{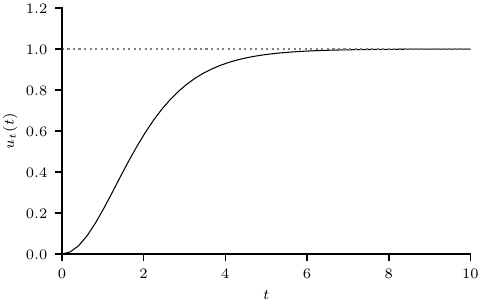}
  \caption{Acceleration of driver}
  \label{fig:kink_driver_accel}
  \end{subfigure}
  \mycaption{Radial velocity profile $u_r(r)$ and acceleration profile
    $u_t(t)$ of the driver \eqref{eq:null_twisting_profile} for
    parameters $u_0 = 0.15$, $r_d = 5$ and $t_r = 2$.}{}
  \label{fig:kink_driver}
\end{figure}

The flux rope is formed by prescribing a slowly accelerating, rotating flow at
the upper $z$-boundary as
\begin{equation}
  \label{eq:null_twisting_profile}
  \vec{u} = u_0 u_r(r) u_t(t) (-y, x, 0)^T,
\end{equation}
where $u_r(r)$ describes the radial profile of the twisting motion in terms of
the radius $r^2 = x^2 + y^2$,
\begin{equation}
  \label{eq:radial_twisting_function}
  u_r(r) = u_{r0}(1 + \tanh(1 - r_d r^2)),
\end{equation}
where $r_d$ controls the radial extent of the driver, $u_{r0}$ is a normalising
factor, and $u_t(t)$ describes the imposed acceleration of the twisting motion,
\begin{equation}
  \label{eq:ramping_up_function}
  u_t(t) = \tanh^2(t/t_r),
\end{equation}
where the parameter $t_r$ controls the time taken to reach the final driver
velocity $u_0$. The functions $u_r(r)$ and $u_t(t)$ are plotted in
Figure~\ref{fig:kink_driver}. At the lower boundary, the flow is in the
opposite direction. This form of driver allows the system to be accelerated
slowly enough that the production of disruptive shocks and fast waves is
minimal. It is unavoidable that some waves are produced during the boundary
acceleration, however these usefully provide a source of noise which eventually
forms a perturbation.

The driver velocity is set to $u_0 = 0.15$, the normalising factor is $u_{r0}
= 2.08$, and setting $r_d = 5$ corresponds to a driver constrained to $r<1$ and
with a peak velocity at $r\approx 0.38$. The ramping time is set to $t_r = 2$,
resulting in an acceleration from $0$ to $u_0$ over approximately $5$ Alfv\'en
times. These driver parameters correspond to a peak rotational period of $T_R
= 15.92$, the length of time taken for one full turn to be injected by a single
driver. Both drivers result in twist being added at a rate of $2\pi$ every
$7.96$ Alfv\'en times. The twist profile across the entire flux tube develops
in such a way that by $t\approx 20$, it is qualitatively similar to those
studied
by~\revcite{\citet{quinnEffectAnisotropicViscosity2020,hoodCoronalHeatingMagnetic2009}}, and \revcite{\citet{barefordShockHeatingNumerical2015}},
however the length of the flux tubes differs significantly. This configuration
produces a $z$-directed tube of increasingly twisted magnetic field that
eventually becomes unstable to both the flute instability and the helical kink
instability.

The problem formulated above is solved numerically using the staggered-grid,
Lagrangian–Eulerian remap code for 3D MHD simulations Lare3D
of~\cite{arberStaggeredGridLagrangian2001} where a  new module for anisotropic
viscosity has been included as detailed
by~\revcite{\citet{quinnKelvinHelmholtzInstabilityCollapse2021}}. The resolution used in
the current work is  $512$ grid points per dimension, comparable to the highest
resolution kink instability studies of~\cite{hoodCoronalHeatingMagnetic2009} or
medium resolution studies of~\cite{barefordShockHeatingNumerical2015}. 

\begin{figure*}
  \centering
    \begin{subfigure}{0.32\textwidth}
      \includegraphics[width=\linewidth]{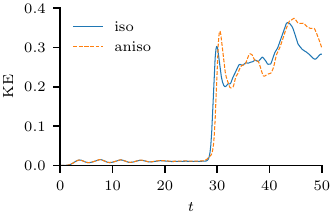}
      \caption{Kinetic Energy}
      \label{fig:kink_ke-4}
    \end{subfigure}
    \hfill
    \begin{subfigure}{0.32\textwidth}
      \includegraphics[width=\linewidth]{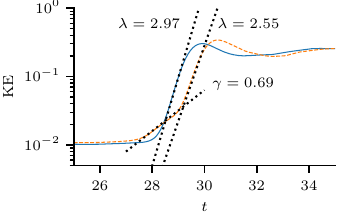}
      \caption{Growth rate estimation}
      \label{fig:kink_ke_log-4}
    \end{subfigure}
    \hfill
    \begin{subfigure}{0.32\textwidth}
      \includegraphics[width=\linewidth]{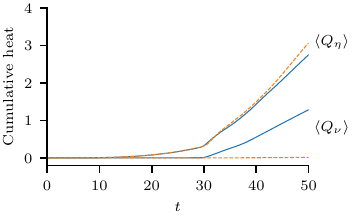}
      \caption{Cumulative heating}
      \label{fig:heating-4}
    \end{subfigure}
  \mycaption{Kinetic energy (a,b) \rev{and cumulative heating (c)} as a function of time showing the development and
  measured growth rates $\gamma$ and $\lambda$ of the flute and kink
  instabilities, respectively.}{Resistivity value is $\eta=10^{-4}$
    and Figure~\ref{fig:kink_ke_log-4} is an enlarged version of
    Figure~\ref{fig:kink_ke-4}. \rev{The cumulative heating $\langle
      Q_{*} \rangle$, where $*$ is either $\nu$ for viscous heating or
      $\eta$ for Ohmic heating, is the respective heating term  
      integrated both over space and from the initial moment up to the moment $t$ in
      time. The viscous heating associated with the flute instability
      (that generated before $t\approx28$) is negligible compared to
      that associated with the kink instability (generated after
      $t\approx28$). While the isotropic 
      model permits greater viscous heating (on the order of two
      orders of magnitude), the anisotropic model enhances Ohmic
      heating.}}
\label{fig:kink_str8_ke-4}%
\end{figure*}

\section{Results}
\label{sec-results}

\rev{In an attempt to extrapolate to coronal resistivity values,} we
focus the attention on two selected pairs of  simulations, one pair
where the background resistivity is set to $\eta=10^{-3}$ and another where
$\eta=10^{-4}$. As in the work of ~\revcite{\citet{quinnEffectAnisotropicViscosity2020}}, only
background resistivity is used. Each pair consists of one simulation using
isotropic viscosity~\eqref{eq:isotropic_viscous_tensor} and another one using
the anisotropic model~\eqref{eq:switching_model}. The value of viscosity is set
to $\nu = 10^{-4}$ in all cases. 

The overall development of both the flute and the kink instabilities is broadly
similar for the two values of resistivity and is described  initially. Similar
simulations were performed with a longer flux tube of length $20$ instead of
the tubes with length $4$ shown here, and the results were found to be
qualitatively similar. Focus is then placed on the detailed description of
instabilities in the $\eta=10^{-4}$ cases, with the aim of comparing the
effects of the two viscosity models. Then further features of the
$\eta=10^{-3}$ cases are summarised.

\subsection{Mechanism and
    general development of instability}

\begin{figure*}
  \centering
    \begin{subfigure}{0.32\textwidth}
      \includegraphics[width=\linewidth]{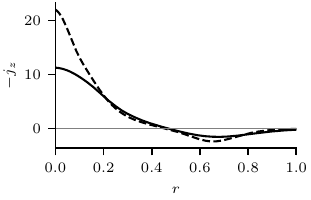}
      \caption{Current}
      \label{fig:current_profiles}
    \end{subfigure}
    \hfill
    \begin{subfigure}{0.32\textwidth}
      \includegraphics[width=\linewidth]{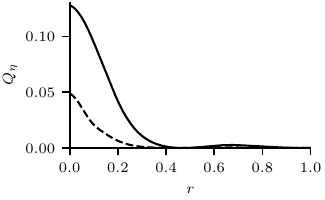}
      \caption{Ohmic heating}
      \label{fig:kink_straight_ohmic_heating_profile}
    \end{subfigure}
    \hfill
    \begin{subfigure}{0.32\textwidth}
      \includegraphics[width=\linewidth]{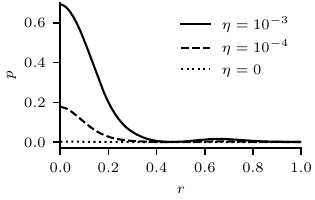}
      \caption{\rev{Gas pressure}}
      \label{fig:pressure_profiles}
    \end{subfigure}
  \mycaption{Gradients in the current density generate \rev{gas} pressure gradients
  through Ohmic heating.}{Axial current density (a), Ohmic heating (b) and
  pressure (c) as functions of the radial distance from the tube axis. All
  plots are from anisotropic cases when $t=20$ through the plane $z=0$. Note
  the sign of the axial current density $j_z$ has been flipped for comparison
  and the Ohmic heating is given by $Q_{\eta} = \eta j^2$. The
  \rev{gas} pressure profile of an additional test-case where $\eta=0$
  is also shown. Line types are   indicated in the legend.}
  \label{fig:pressure_and_heating}
\end{figure*}

Initially and in all cases computed, the twisting at the upper and lower
boundaries gives rise to a pair of torsional Alfv\'en waves which proceed to
travel along the tube from the upper and lower boundaries to their respective
opposite boundaries. The interaction between these waves produces an
oscillating pattern in the kinetic energy with a period of approximately $4$
Alfv\'en times, equal to the time taken for an Alfv\'en wave to traverse the
entire length of the domain as visible early in Figure~\ref{fig:kink_ke-4}. 

As the field continues to be twisted, currents form, due to the local shear in
the magnetic field, and heat the plasma through Ohmic dissipation. Due to the
radial form of the driver, the magnitude of the current density is greatest at
the axis of the tube, then decreases radially outwards as seen in
Figure~\ref{fig:current_profiles}. The orientation of the twisting produces
a current flowing in the negative $z$-direction for $r\lessapprox0.5$. At $r
\approx 0.5$ (corresponding to the radius of peak driving velocity) the current
switches orientation and is in the positive $z$-direction in a shell where
$0.5\lessapprox r \lessapprox 0.8$. This form of a twisted field with an inner
core of current in one direction surrounded by a shell of oppositely-directed
current is similar to the current configuration arising due to the field
prescribed by~\revcite{\citet{quinnEffectAnisotropicViscosity2020}}. 

This current profile is reflected in the radial Ohmic heating profile
(Figure~\ref{fig:kink_straight_ohmic_heating_profile}) and, consequently, in
the radial \rev{gas} pressure profile (Figure~\ref{fig:pressure_profiles}). The highly
pressurised core extends to $r\approx 0.2$--$0.4$ (depending on the value of
$\eta$) before increasing slightly around $r\approx 0.7$. The secondary bump in
\rev{gas} pressure is due to the outer shell of current. The \rev{gas}
pressure gradient near the
axis provides the outwardly directed \rev{gas} pressure force which competes against the
binding action of the magnetic tension and this provides the mechanism of flute
instability excitation.  The magnitude of the \rev{gas} pressure gradient depends
strongly on the value of resistivity $\eta$, with lower values producing
shallower gradients which (as shall be seen) are more stable to the flute
instability. Indeed, when $\eta=0$, the radial \rev{gas} pressure profile is nearly flat
and the tube stable to the flute instability.

In all cases unstable to the flute instability, it occurs between $t=20$ and
$t=30$. The continued driving at the boundaries eventually injects enough twist
that the tube also becomes unstable to the kink instability. The kink initially
develops linearly alongside or shortly after the flute instability and then
\rev{erupts during its nonlinear phase}, dominating the dynamics and disrupting
the   flute instability. The onset and the competition of the two instabilities
is strongly affected by the value of $\eta$ and the viscosity model used. 

\subsection{Instabilities at resistivity $\eta=10^{-4}$}

\begin{figure*}
  \centering
    \begin{subfigure}{0.32\textwidth}
      \includegraphics[width=\linewidth]{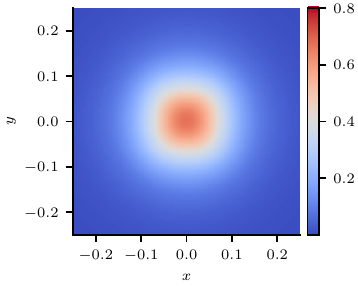}
      \caption{$t=26$}
      \label{fig:swi-4_pressure_13}
    \end{subfigure}
    \hfill
    \begin{subfigure}{0.32\textwidth}
      \includegraphics[width=\linewidth]{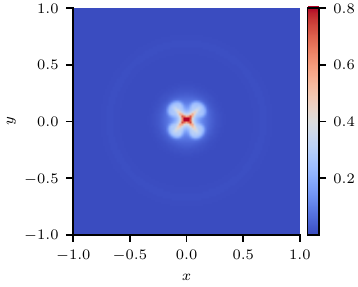}
      \caption{$t=28$}
      \label{fig:swi-4_pressure_14}
    \end{subfigure}
    \hfill
    \begin{subfigure}{0.32\textwidth}
      \includegraphics[width=\linewidth]{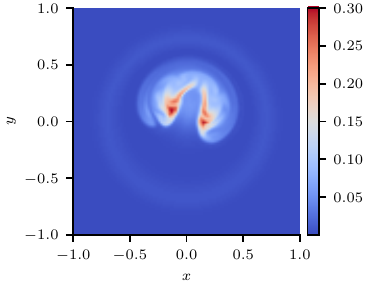}
      \caption{$t=30$}
      \label{fig:swi-4_pressure_15}
    \end{subfigure}
\mycaption{\rev{Gas pressure} profiles during the development of the flute and kink
instabilities.}{Shown are density plots of \rev{gas} pressure at $z=0$ with $\eta
= 10^{-4}$ and for the anisotropic viscosity model. Note the difference in
colour scale in Figure~\ref{fig:swi-4_pressure_15}. The development in the
isotropic case is similar.}
\label{fig:kink_pressure_slices-4}%
\end{figure*}

We now describe the evolution and competition of flute and kink instabilities
in case of resistivity $\eta=10^{-4}$. Figure~\ref{fig:kink_pressure_slices-4}
shows the \rev{gas} pressure profile of the anisotropic viscosity case
\eqref{eq:switching_model} at time moments $t=26$, $28$ and $30$ and at $z=0$.
\rev{At $t= 26$ the plasma begins to bulge out diagonally} from the
high-pressure core displaying an azimuthal wavenumber $m=4$ as seen in
Figure~\ref{fig:swi-4_pressure_13} \rev{and indicating the presence of the
flute instability}. As  the 
bulges move radially outwards into lower \rev{gas} pressure regions they expand and
accelerate, resulting in the entire \rev{gas} pressure structure \rev{\newold{}{appearing}} taking the
shape of a four-leaf clover (Figure~\ref{fig:swi-4_pressure_14}). By $t=30$ the
kink instability has disrupted the flute instability and is developing
nonlinearly as evident in Figure~\ref{fig:swi-4_pressure_15}. As is typical of
nonlinear kink development, the tube continues to release its stored potential
energy in the form of kinetic energy and heat and the contained plasma becomes
highly mixed. In the  isotropic viscosity case which will not be illustrated by
a separate figure, the flute instability is present but \rev{its
 growth is damped relative to the anisotropic case}, and it is quickly
outcompeted by the kink instability which  dominates the dynamics. 

\begin{figure}
  \centering
    \begin{subfigure}{0.49\textwidth}
      \includegraphics[width=\linewidth]{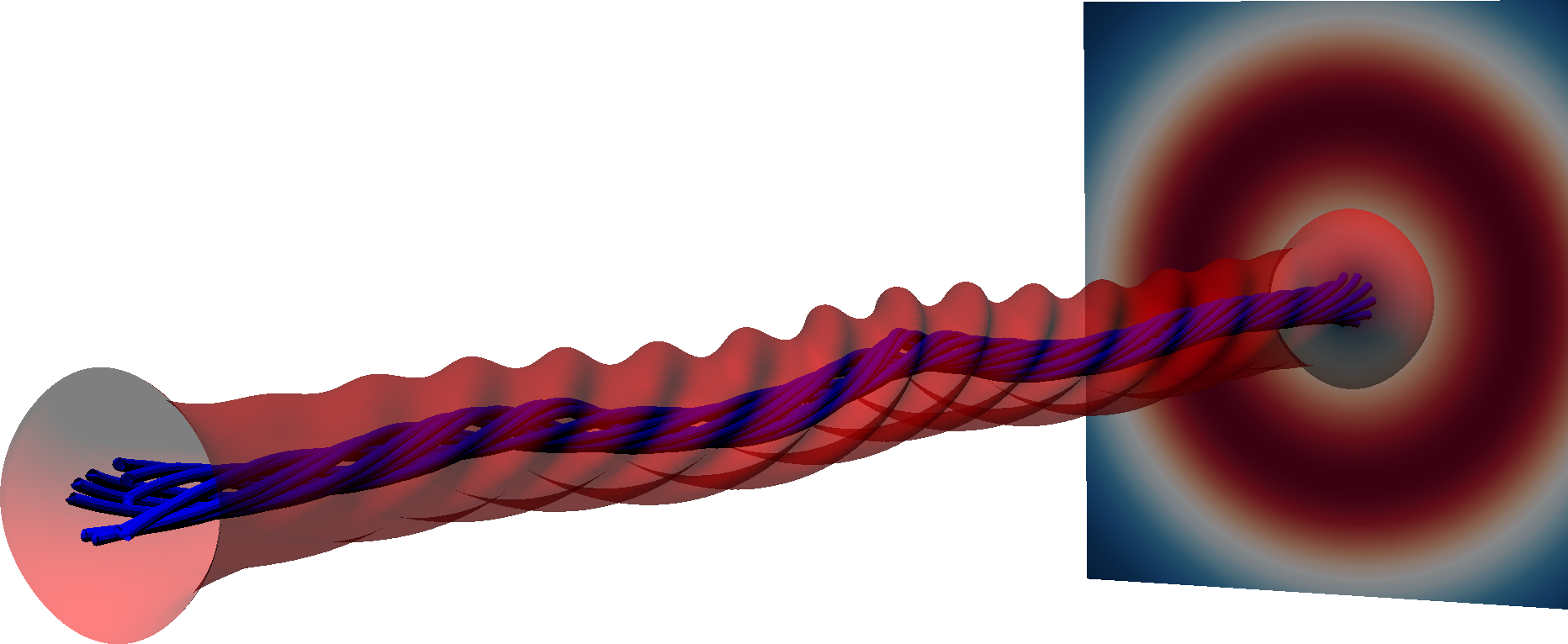}
      \caption{Isotropic}
      \label{fig:field_line_plots_iso}
    \end{subfigure}
    \hfill
    \begin{subfigure}{0.49\textwidth}
      \includegraphics[width=\linewidth]{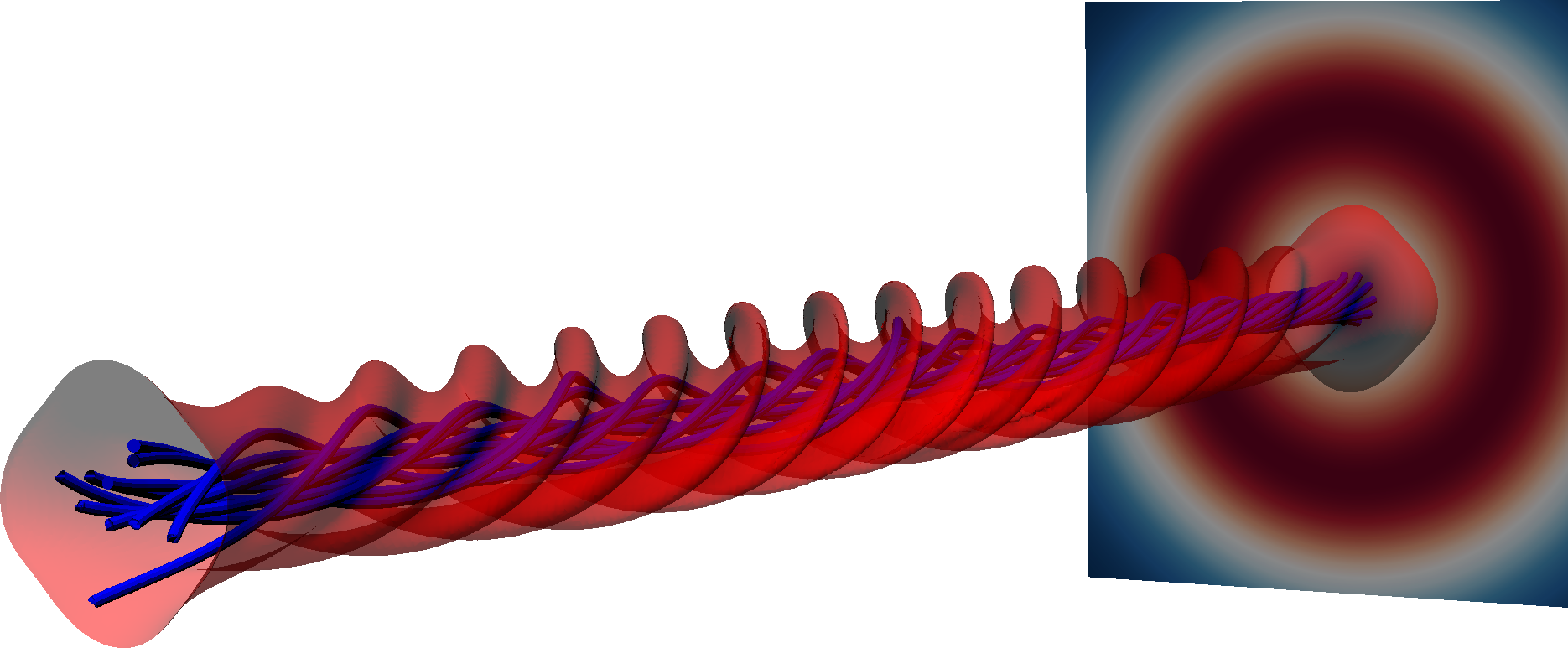}
      \caption{Anisotropic}
      \label{fig:field_line_plots_swi}
    \end{subfigure}
\mycaption{Simultaneous development of flute and kink instabilities in the
isotropic and anisotropic cases illustrated by field lines and
\rev{gas} pressure
contours.}{Field lines and contours of \rev{gas} pressure (where $p=0.3$) are plotted at
$t=28$. Also shown is the velocity driver $u_r(\sqrt{x^2+y^2})$ at $z=2$. The
flute instability grows in both cases, though faster in the anisotropic case.
The initial stages of the kink instability can also be observed in the field
lines of the isotropic case in subfigure~\ref{fig:field_line_plots_iso}.}
\label{fig:kink_field_line_plots}%
\end{figure}

Figure~\ref{fig:kink_field_line_plots} shows the effect the viscosity models
have on the initial stages of the flute and kink instabilities in 3D. While the
flute instability is observed in both cases, \rev{it is more
  pronounced in the anisotropic case, where it} appears to disrupt the inner
core of field lines and, as will be discussed further below, slows the growth
of the kink instability. In the isotropic case the \rev{growth of the flute
instability is damped relative to the anisotropic case} to the extent that the
kink instability grows uninhibited and quickly disrupts the fluting. 

Despite the flute instability appearing in the isotropic case
(Figure~\ref{fig:field_line_plots_iso}), only in the anisotropic case
can \rev{phases of linear growth of} the flute and kink instabilities be seen in the kinetic energy
profile as shown in Figure~\ref{fig:kink_ke_log-4}. Here, the growth
rates of the two instabilities are found to be $\gamma = 0.69$ for the flute
and $\lambda = 2.55$ for the kink. \rev{The apparent phases of linear
growth as measured from the kinetic energy time series, start} at approximately $t=27$ for
the flute instability and $t=29.5$ for the kink. In the isotropic case, the
growth rate of the kink, $\lambda = 2.97$, is larger than in the anisotropic
case, while the kinetic energy profile shows no evidence of flute instability growth. 

The faster growth of the kink compared to that measured by~\revcite{\citet{quinnEffectAnisotropicViscosity2020}} is attributed to the relative
aspect ratios of the flux tubes. The tube prescribed by~\revcite{\citet{quinnEffectAnisotropicViscosity2020}} has an aspect ratio of
approximately $20$ compared to the tube studied here which has an aspect ratio
of approximately $4$. While the total twist is similar in both tubes (after the
drivers have injected twist up to $t\approx20$) the small aspect ratio results
in more turns per unit length, leading to a faster growing instability. 

\begin{figure}
  \centering
    \begin{subfigure}{0.49\textwidth}
      \includegraphics[width=\linewidth]{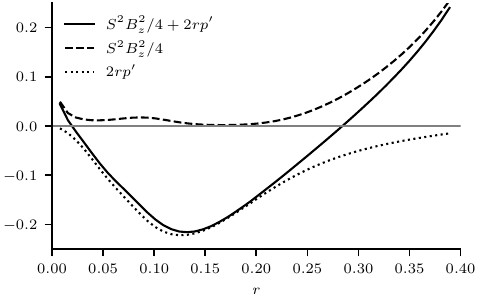}
      \caption{Suydam condition}
      \label{fig:suydam_condition_4}
    \end{subfigure}
    \hfill
    \begin{subfigure}{0.49\textwidth}
      \includegraphics[width=\linewidth]{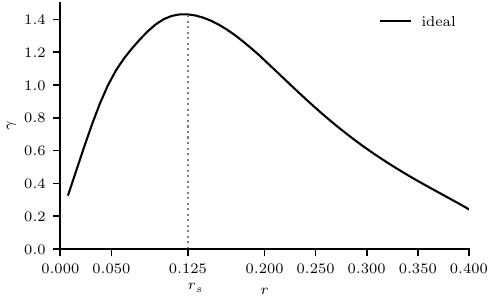}
      \caption{Linear growth rates}
      \label{fig:growth_rate_4}
    \end{subfigure}
\mycaption{Stability and linear growth rate of the flute instability.}{In panel
\ref{fig:suydam_condition_4}, Suydam's stability criterion
\eqref{eq:suydams_criterion} and its contributing terms  are plotted and in
panel \ref{fig:growth_rate_4} the predicted linear growth rates for the
ideal~\eqref{eq:fluting_growth_rate2} flute instability are plotted. Both plots
are produced at $t=20$ for $\eta=10^{-4}$ and using the anisotropic model. The
location of the \rev{peak linear ideal growth rate $r_s$} is also shown.}
\label{fig:stability_and_growth}%
\end{figure}

Prior to the onset of either instability, the flux tube is found to be linearly
unstable to perturbations of the form~\eqref{eq:kink_perturbation} at $t=20$
via Suydam's criterion~\eqref{eq:suydams_criterion} as shown in
Figure~\ref{fig:suydam_condition_4}. The criterion represents a
balance between destabilising pressure gradients and stabilising
magnetic shear and in this case, the shear is so small and the
pressure gradient so large that the tube is unstable over a wide range
of radii, for $ 0.02 \lessapprox r \lessapprox 0.29$.
\rev{Using equation~\eqref{eq:fluting_growth_rate2}, the} linear
fluting growth rate $\gamma$ is plotted as a function of $r$ at
$t=20$ in Figure~\ref{fig:growth_rate_4}. \rev{At any fixed moment,
the radial dependence of the flute instability growth rate, $\gamma(r)$, is a concave
function and peaks at a certain radius that we denote by $r_s$ in
Figure~\ref{fig:growth_rate_4} and Table~\ref{tab:kink_fluting_params}.}

The location \rev{$r_s$} of the peak \rev{fluting growth rate} aligns
well with the location of the resonant surface where the observed
perturbation appears to grow 
(Figure~\ref{fig:swi-4_pressure_13}) and an estimate of the linear growth rate
can be found by averaging $\gamma$ over $r$, giving a theoretical growth rate of $0.88$.

\begin{figure}
  \centering
    \begin{subfigure}{0.49\textwidth}
      \includegraphics[width=\linewidth]{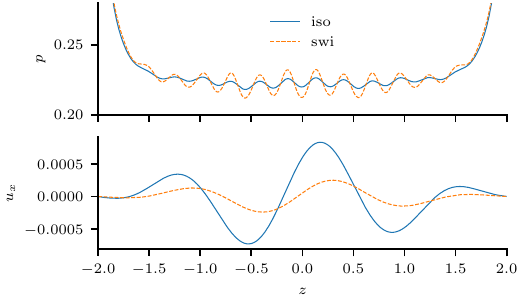}
      \caption{\rev{Selected flow and pressure profiles.}}
      \label{fig:pressure_pert_4}
    \end{subfigure}
    \hfill
    \begin{subfigure}{0.49\textwidth}
      \includegraphics[width=\linewidth]{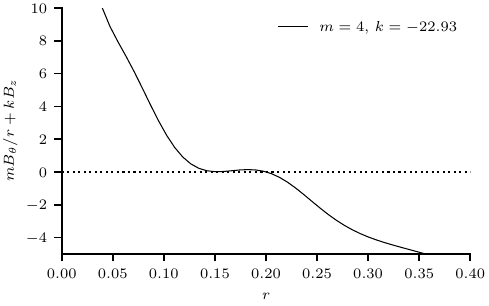}
      \caption{Resonance function.}
      \label{fig:resonant_surface_4}
    \end{subfigure}
  \mycaption{\rev{\newold{Selected flow and pressure profiles and the 
resonance function defined by equation
\eqref{eq:resonant_surface}.}{Perturbations corresponding to the flute
        and kink instabilities and the spatial radial distribution of the associated resonance
function.}}}{\rev{(a) Gas pressure} and velocity \rev{profiles in $z$}
    \rev{at the fixed point $(r, \theta) = (0.101, 0)$} \rev{assuming the form of the most
      unstable modes}. (b) The resonance surface $m B_{\theta}(r)/r
+ kB_z(r)$ as a function of $r$ using the observed fluting perturbation
wavenumbers. All plots are  snapshots at $t=26$ where $\eta=10^{-4}$ and the
viscosity model is anisotropic.} 
\label{fig:k_and_resonance}%
\end{figure}

The \rev{flow and pressure profiles in the axial direction $z$} at
$t=26$ are shown in Figure~\ref{fig:pressure_pert_4} \rev{and at this
moment they assume the form of a superposition of the unstable modes
with the largest amplitude. In particular, the} fluting perturbation
  is most easily observed in the pressure \rev{profile and} \rev{\newold{}{ plotted
  as a function of $z$ following a line through the point $(r, \theta) = (0.101, 0)$.}} \rev{the} kink
instability is best revealed \rev{by either of the $x$- or
the $y$-component of} velocity which can serve as proxies for the radial
velocity through the axis. Comparing the magnitudes of the \rev{profiles} at
this time suggests the flute instability is close to transitioning to its
nonlinear phase while the kink instability is still very much in its linear
phase.

The value of $k$ for each \rev{instability mode} is calculated as $k
= 2\pi/\tilde{\lambda}$ where $\tilde{\lambda}$ is the wavelength of the
perturbation, measured as the difference between the two peaks closest
to $z=0$ \rev{in Figure~\ref{fig:pressure_pert_4}}
(thus minimising the influence of line-tying on the measurement). This gives
a value of $k_{\text{flute}}=22.93$ and $k_{\text{kink}}=4.57$ for the
anisotropic model and $k_{\text{flute}}=22.30$ and $k_{\text{kink}}=4.41$.
Using these values, it is observed that the fluting mode resonates with
the field, that is $m B_{\theta}(r)/r + kB_z(r) \approx 0$, over a range of
$0.15 \lesssim r \lesssim 0.225$ (Figure~\ref{fig:resonant_surface_4}). This is
in close agreement with the predicted radius of peak linear \rev{flute growth
$r_s$} seen in Figure~\ref{fig:growth_rate_4}.

Comparing the effect of the viscous models on the modes, \rev{in the
isotropic case the fluting mode is damped}, while in the anisotropic
case the kink mode is diminished, explaining why the flute instability
appears more readily in the anisotropic case (Figure~\ref{fig:kink_ke-4}).

\subsection{Instabilities at resistivity $\eta=10^{-3}$}

\begin{figure*}
  \centering
    \begin{subfigure}{0.32\textwidth}
      \includegraphics[width=\linewidth]{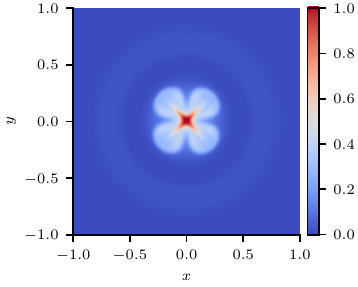}
      \caption{$t=24$}
      \label{fig:swi-3_pressure_12}
    \end{subfigure}
    \hfill
    \begin{subfigure}{0.32\textwidth}
      \includegraphics[width=\linewidth]{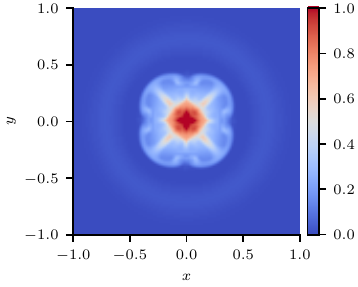}
      \caption{$t=28$}
      \label{fig:swi-3_pressure_14}
    \end{subfigure}
    \hfill
    \begin{subfigure}{0.32\textwidth}
      \includegraphics[width=\linewidth]{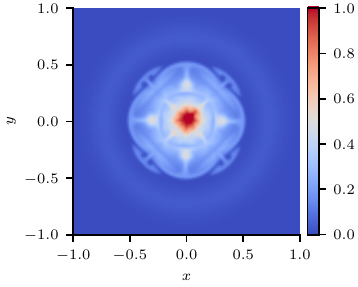}
      \caption{$t=30$}
      \label{fig:swi-3_pressure_15}
    \end{subfigure}
    \hfill
    \begin{subfigure}{0.32\textwidth}
      \includegraphics[width=\linewidth]{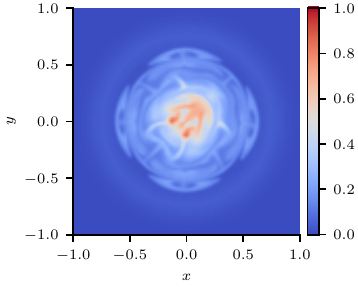}
      \caption{$t=32$}
      \label{fig:swi-3_pressure_16}
    \end{subfigure}
    \hfill
    \begin{subfigure}{0.32\textwidth}
      \includegraphics[width=\linewidth]{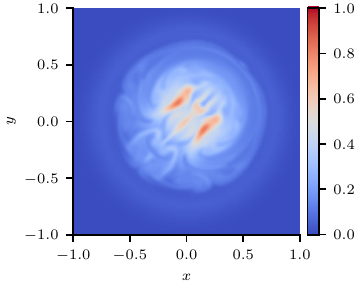}
      \caption{$t=34$}
      \label{fig:swi-3_pressure_17}
    \end{subfigure}
    \hfill
    \begin{subfigure}{0.32\textwidth}
      \includegraphics[width=\linewidth]{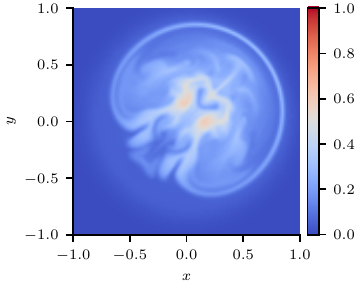}
      \caption{$t=36$}
      \label{fig:swi-3_pressure_18}
    \end{subfigure}
\mycaption{\rev{Gas pressure} profiles at $z=0$ during the development of the flute and
kink instabilities in the higher resistivity anisotropic case.}{The viscosity
model is anisotropic and $\eta = 10^{-3}$. In contrast to the case of
$\eta=10^{-4}$, the nonlinear development of the flute instability has time to
mix the interior of the flux tube before the onset of the kink instability, the
growth of which is \rev{likely} affected by mixing of the plasma.}
\label{fig:kink_pressure_slices-3}%
\end{figure*}

Figure \ref{fig:kink_pressure_slices-3} shows a prolonged development of the
flute instability and a slow nonlinear development of the kink instability at
the higher resistivity value $\eta=10^{-3}$ in the case of anisotropic
viscosity. Due to the enhanced Ohmic heating at $\eta=10^{-3}$, the \rev{gas} pressure
gradient is substantially stronger than at $\eta=10^{-4}$ and the flute
instability is excited much earlier. Compared to the $\eta=10^{-4}$ cases, the
instability occurs further from the axis, at $r\approx0.16$, and the larger
\rev{gas} pressure gradient drives the bulges in profile further from the axis during the
nonlinear phase (Figure~\ref{fig:swi-3_pressure_12}). These bulges continue to
extend outwards and mix the plasma as they develop. The kink instability can be
observed displacing the axis of the tube diagonally upwards and to the right in
Figure~\ref{fig:swi-3_pressure_15}. At this time in the $\eta=10^{-4}$ cases,
the nonlinear development of the kink was at a later stage of its development
(Figure~\ref{fig:swi-4_pressure_15}). The development of the kink then proceeds
slowly as it moves the axis of the tube further through the mixed
region to eventually begin the reconnection process with the outer
region of field that is typical of the instability in this kind of
flux tube as was observed by
\revcite{\citet{quinnEffectAnisotropicViscosity2020}}.  

\begin{figure*}
  \centering
    \begin{subfigure}{0.32\textwidth}
      \includegraphics[width=\linewidth]{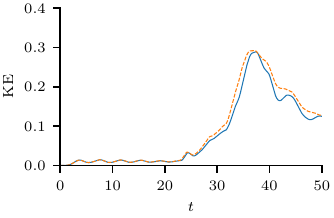}
      \caption{Kinetic Energy}
      \label{fig:kink_ke-3}
    \end{subfigure}
    \hfill
    \begin{subfigure}{0.32\textwidth}
      \includegraphics[width=\linewidth]{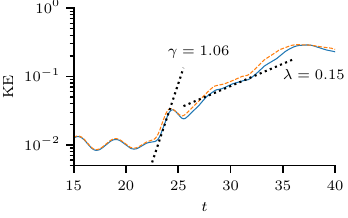}
      \caption{Growth rate estimation}
      \label{fig:kink_ke_log-3}
    \end{subfigure}
    \hfill
    \begin{subfigure}{0.32\textwidth}
      \includegraphics[width=\linewidth]{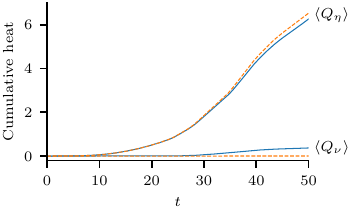}
      \caption{Cumulative heat}
      \label{fig:heating_r-3}
    \end{subfigure}
\mycaption{Kinetic energy \rev{and heating} as a function of time in the cases where
$\eta=10^{-3}$.}{The results from both viscosity models are shown. The flute
instability appears earlier than where $\eta=10^{-4}$ and the growth rate of
the kink instability is decreased.
\rev{The cumulative heating $\langle
      Q_{*} \rangle$, where $*$ is either $\nu$ for viscous heating or
      $\eta$ for Ohmic heating, is the respective heating term  
      integrated both over space and from the initial moment up to the moment $t$ in
      time.         While the heat generated via viscous heating is
      orders of magnitude lower when using anisotropic viscosity,
      Ohmic heating is enhanced by the use of the anisotropic model.}} 
\label{fig:kink_str8_ke-3}%
\end{figure*}

It is evident from the kinetic energy profile that the flute instability
develops much earlier than in the $\eta=10^{-4}$ cases and grows at an
increased rate of $\gamma = 1.06$ (Figure~\ref{fig:kink_ke_log-3}). The kink
instability grows at a rate of $\lambda \approx 0.15$, much slower than that
observed in the $\eta=10^{-4}$ cases, and much lower than the flute
instability. \rev{The time period between $t\approx28$ and $t\approx32$ is
identified as the linear stage of the kink instability by inspecting the
development shown in Figure~\ref{fig:swi-3_pressure_12}. This is broadly
consistent with the tube surpassing a critical twist of
$2.59\pi$~\citep{hood1981,Torok2003} between $t=30$ and $t=35$.} One key
observation is that, despite the early and disruptive growth of the flute
instability, the kink instability still generates the bulk of the kinetic
energy (Figure~\ref{fig:kink_ke-3}).

Due to the influence of the drivers on the kinetic energy, the fluting growth
rate is difficult to estimate from the kinetic energy profile as accurately as
in the $\eta=10^{-4}$ cases. Since the kink instability occurs after the
development of the fluting, its growth rate is similarly difficult to gauge.
Nevertheless, it is clear that the flute instability grows at a rate of the
same order as that in the $\eta=10^{-4}$ cases. It is also apparent that the
kink instability grows much slower in the $\eta=10^{-3}$ cases.

\begin{table*}
  \caption{
Quantitative differences in the observed perturbations between results
\rev{with} different \rev{viscosity models and} resistivity values
$\eta$. Measurement times are listed in the main text.}  
\centering
\begin{tabular}{ccccc}
&\multicolumn{2}{c}{$\eta=10^{-4}$}& \multicolumn{2}{c}{$\eta=10^{-3}$} \\
&
Anisotropic &
\rev{Isotropic} &
Anisotropic &
\rev{Isotropic} \\
\midrule
Theoretical average ideal linear growth rate of flute $\gamma$
& 0.88 
& \rev{0.88} 
& 1.73 
& \rev{1.73} \\
Observed growth rate of flute $\gamma$
& 0.69 
& \rev{--} 
& 1.06 
& \rev{1.06}\\
Observed growth rate of kink $\lambda$
& 2.55 
& \rev{2.97} 
& 0.15 
& \rev{0.15}\\
\midrule
Theoretical radius $r_s$ of peak ideal flute growth rate
& 0.125 
& \rev{0.125} 
& 0.163 
& \rev{0.163}\\
\midrule
\rev{Predicted axial wave number $k_{\text{flute}}$}
& \rev{23.74}
& \rev{23.52}
& \rev{17.15}
& \rev{17.60}\\
Observed axial wave number $k_{\text{flute}}$
& \rev{$22.93$}
& \rev{$22.30$}
& \rev{$16.05$}
& \rev{$16.05$}\\
Observed axial wave number $k_{\text{kink}}$
& {4.57}
& \rev{4.41}
& {3.44}
& \rev{3.49}\\
\midrule
\rev{Cumulative viscous heat at $t=50$}
& \rev{$1.64 \times 10^{-2}$}
& \rev{1.28}
& \rev{$2.89 \times 10^{-3}$}
& \rev{0.370} \\
\rev{Cumulative Ohmic heat at $t=50$}
& \rev{3.06}
& \rev{2.75} 
& \rev{6.54} 
& \rev{6.27} \\
\end{tabular}
\label{tab:kink_fluting_params}
\end{table*}

Table~\ref{tab:kink_fluting_params} summarises the quantitative differences
between the results for \rev{the two models
  of viscosity and} the two values of the resistivity. \rev{The
  theoretical average growth rate is computed as the mean across the
  radius of the ideal estimate~\eqref{eq:fluting_growth_rate2} and is
  in good agreement with the observed rate in each case, particularly
  in the less resistive case $\eta=10^{-4}$ which better represents
  ideal plasma. The
  discrepancy between predicted and observed growth rates is due,
  in part, to the growth rate estimate~\eqref{eq:fluting_growth_rate2}
  being derived under the assumption of asymptotically large values of
  $m \gg 1$, while the observed mode has a finite value of
  $m=4$. Despite this, the predicted growth rate is of a similar 
  magnitude to the observed rate. The location $r_s$ of the 
peak growth rate provides a prediction of where the instability will initially
grow. This radius is used in conjunction with the resonance
equation~\eqref{eq:resonant_surface}, with $m=4$, to predict the axial
wavenumber of the mode with the greatest linear growth, i.e. the
mode most likely to be observed. Again, these are in good agreement
with the observed fluting wavenumbers, which are measured at times just prior
to the accelerated development of the flute instability, that is at $t=22$ when
$\eta=10^{-3}$ and $t=26$ when $\eta = 10^{-4}$. The kink wavenumber is
measured at $t=26$ in both cases. Overall, the agreement between
predicted and observed growth rates and mode wavenumbers
allows us to conclude that the observed instability is the ideal flute
instability and that expression~\eqref{eq:fluting_growth_rate2}
can be effectively applied to estimate the growth rate of the flute
instability within coronal loops.}  

\rev{Also listed are estimates for
the cumulative heat generated via viscous and Ohmic heating during the
simulations. As is also found in previous studies of viscous heating in
kink instabilities~\citep{quinnEffectAnisotropicViscosity2020}, anisotropic
viscous heating is approximately two orders of magnitude lower than isotropic
and the use of anisotropic viscosity enhances Ohmic heating.}

\section{Discussion}
\label{sec-discussion}

\rev{Perturbing a magnetic flux tube by dynamic twisting of the flow
at the cylinder bases leads to excitation of the flute instability in
addition to the well studied kink instability.}
\rev{Our aim in performing the reported numerical experiments was to explore
the flute instability in a dynamically twisted coronal flux tube, specifically
focussing on the effect of anisotropic viscosity on the development of the
instability. In addition, we wish to understand the effect the instability has
on the proceeding kink instability, the effect on the overall heating generated
through viscous and Ohmic dissipation, and the effect that varying resistivity
has on the development of the flute. Our findings are discussed below.}

\rev{We have found evidence of the flute instability using both models of
viscosity, however isotropic viscosity damps the initial growth such that it
does not develop beyond its linear phase before the faster growing kink
instability disrupts the flux tube and halts the development of the flute.
Given that most numerical studies of the kink instability employ isotropic or
shock viscosity, this is likely why the flute instability has not been
previously reported.}

\rev{Counter-intuitively, the growth rate of the kink instability is lower in the
weakly dissipative anisotropic cases, compared to the strongly dissipative
isotropic cases as one would expect the kink instability to grow more
quickly (or at least be unaffected) when using anisotropic viscosity. Indeed,
the simulations reported by~\citet{quinnEffectAnisotropicViscosity2020} display
this behaviour, where the kink instability grows \newold{}{(marginally)} faster in the
anisotropic cases. We speculate that it is the presence of the flute modes which
negatively affects the growth of the kink instability.} {It seems unlikely that
in the linear regime the flute and kink modes are able to
directly couple, given that the kink instability generally presents at the axis
of a flux tube and the flute at some resonant surface away from the
axis.} \rev{We believe that, instead of a direct coupling, the linear kink instability is
disturbed by the more complex magnetic field configuration that arises due to
the mixing caused by the nonlinear development of the flute modes. The
complexity in the field can be seen by comparing
figures~\ref{fig:field_line_plots_iso} and~\ref{fig:field_line_plots_swi}.

Beyond the effect on the growth of the two observed instabilities, the two
viscosity models greatly affect both the viscous heating and Ohmic heating
rates illustrated in Figures \ref{fig:heating-4}
and~\ref{fig:heating_r-3}. Anisotropic viscosity is naturally less
dissipative than isotropic 
viscosity and generates approximately two orders of magnitude less total
viscous heat than isotropic viscosity. This is somewhat offset by anisotropic
viscosity permitting greater kinetic energy release and enhanced mixing, which
in turn enhances Ohmic dissipation of heat through the generation of
strong localised current sheets. The overall effect is that more heat is generated when
the viscosity is anisotropic, similar to what has been observed in previous
work~\citep{quinnEffectAnisotropicViscosity2020}. In the context of
coronal heating, this is encouraging: the use of a less dissipative viscosity
model actually results in greater overall heating. How this finding generalises
to more realistic coronal resistivities, and whether it holds true for
other coronal instabilities, should be the subject of further investigations.}

\rev{It is difficult to distinguish the effect of the flute instability on the
viscous or Ohmic heating from that of the viscosity itself, particularly since
the flute instability is quickly disrupted by the kink instability. However, it
can be concluded from the plots of cumulative heat (Figures~\ref{fig:heating-4}
and~\ref{fig:heating_r-3}) that the bulk of the viscous and Ohmic heat is
generated in the nonlinear phase of the kink instability. There is
additionally some non-negligible Ohmic heating generated prior to the
onset of the kink, however this is attributed to the large-scale currents
associated with the twist in the field, rather than any currents created by the
flute instability. This leads us to conclude that the flute instability itself
has little direct impact on coronal heating, but instead can affect the heating
rate by slowing the development of the kink instability.}

It is likely that the $m=4$ azimuthal mode is excited due to influences
from the boundaries in the Cartesian box, for example through the interaction
of reflected fast waves generated in part by the driver. Performing a similar
experiment in a cylindrical numerical domain, or prescribing a variety of
perturbations in the Cartesian domain may reveal other, faster growing modes.
The modes may also be influenced by nonlinear coupling between the $m>1$ and
$m=1$ modes, as is found in the study of kink and flute
oscillations~\citep{terradasEffectMagneticTwist2018,rudermanNonlinearGenerationFluting2017}. 

As the current distribution, which develops as the flux tube is twisted, is
similar to that found in the initial flux tube configuration
of~\cite{quinnEffectAnisotropicViscosity2020}, the question arises why the
fluting instability is not observed in the latter. Although the current
distribution (and thus heating and pressure distributions) in the tubes
of~\cite{quinnEffectAnisotropicViscosity2020} may support the flute
instability, the tube is initially perturbed with a motion close to an unstable
eigenmode of the kink instability, resulting in the instability
growing immediately from $t=0$. In contrast, in the tubes studied here, such a perturbation must grow
from numerical noise, allowing a secondary, fluting modes perturbation to also develop
and become significant enough to observe. \rev{As an explorative alternative to
prescriptive perturbations, we recommend the use of numerical noise in the
study of coronal instabilities.}

Since the main driver of the flute instability is the \rev{gas} pressure gradient
generated through Ohmic heating, it is prudent to ask if the same
\rev{gas} pressure
gradient could be generated using physical coronal values of the resistivity,
which are estimated to be approximately
$\eta=10^{-8}$~\citep{craigAnisotropicViscousDissipation2009a}, and are thus
much smaller than those studied here. Additionally, the simulations presented
here do not incorporate radiation or thermal conduction, two processes which
would remove energy (and hence reduce \rev{gas} pressure) from high-pressure regions in
a coronal loop and thus could prevent meeting the required conditions for the
growth of a flute instability. Indeed, at $\eta=10^{-4}$ the flute instability
was more quickly outcompeted by the kink instability and appeared to have
little impact on the resultant dynamics, which mirror those of other kink
instability studies~\citep{hoodCoronalHeatingMagnetic2009}. This suggests
that even lower values of resistivity would result in flux tubes without any
significant flute instability, at least for this form of driver and mechanism
of \rev{gas} pressure generation. Regardless, coronal loops with
strong radial \rev{gas} pressure
gradients have been observed~\citep{foukalTemperatureStructurePressure1975},
and such loops may be unstable to the flute instability. Modelling of
a prescribed flute-unstable flux tube, as opposed to the dynamically stressed
loop investigated here, would provide a useful comparison to observations,
however it may be difficult to prescribe a tube which is not also susceptible
to kinking. Linear stability analyses of this kind of flux tube (a dynamically
created zero total axial current tube) focus on the kink
instability~\citep{browningSolarCoronalHeating2003b} so do not provide much
insight into the potential for fluting without a kink. 

While our results show that a flux tube can be unstable to the flute
instability and yet the faster growing kink instability can quickly dominate
when the \rev{gas} pressure gradient is small enough, the opposite case is also observed.
A faster growing flute instability appears to slow the growth of the kink
instability although, importantly, it does not fully disrupt the development of
the kink. Understanding the balance between the nonlinear growth rates of the
two instabilities is important for prediction of whether the flute instability
may be found at all in the real solar corona, or whether its realistic growth
rate is too slow compared to that of the kink instability. 

\section{Conclusion}
\label{sec-conclusions}

This paper details the nonlinear development of two ideal instabilities, the
kink and the flute instabilities, both of which develop naturally in the course
of twisting an initially straight magnetic flux tube. This provides a different
approach to that employed in the simulations performed in the earlier
study by ~\revcite{\citet{quinnEffectAnisotropicViscosity2020}} in that the instabilities are
not excited by any prescribed perturbations but, instead, the field is
dynamically driven into an unstable state and the perturbations provided by
noise in the system. Not only is the kink instability excited due to the twist
in the field, but nearly simultaneously a pressure-driven flute instability can
also be excited in unstable pressure gradients generated by Ohmic heating.
Simulations were performed with two values of resistivity, $\eta=10^{-3}$ and
$10^{-4}$, and for two forms of viscosity, isotropic and anisotropic. The
results prove an initial and important first step towards understanding
nonlinear flute instabilities in the solar corona.  

It has been shown that the flute instability can be quickly dominated by the
kink instability if the kink grows substantially faster than the flute.
However, if the flute has time to develop nonlinearly, it mixes the plasma
within the flux tube, generating small scale current sheets and releasing some
magnetic energy. The overall effect of this mixing is to slow the growth of the
kink instability. The slowed growth of the kink does not appear to
significantly impact the kinetic energy released during its evolution, only the
time over which it is released. 

These numerical experiments have provided evidence that the flute instability
can occur in twisted magnetic flux ropes and grow at similar rates to the kink
instability. Further estimation of the relative growth rates in more realistic
coronal loop setups is required to fully understand if the flute instability
plays a pertinent role in coronal loop physics.

\section*{Acknowledgements}

We would like to thank David MacTaggart for his input on the thesis chapter on
which this paper is based. JQ was funded via an EPSRC studentship: EPSRC DTG EP/N509668/1.

Results were obtained using the ARCHIE-WeSt High
Performance Computer (\url{www.archie-west.ac.uk}) based at the University of
Strathclyde. 

The authors acknowledge the use of the UCL Myriad High Performance Computing Facility (Myriad@UCL), and associated support services, in the completion of this work

This work used the DiRAC Extreme Scaling service at the University of Edinburgh, operated by the Edinburgh Parallel Computing Centre on behalf of the STFC DiRAC HPC Facility (www.dirac.ac.uk), specifically Thematic Project allocation ACTP245. This equipment was funded by BEIS capital funding via STFC capital grant ST/R00238X/1 and STFC DiRAC Operations grant ST/R001006/1. DiRAC is part of the National e-Infrastructure.

\bibliographystyle{mnras}
\bibliography{paper}

\appendix

\section{Associated software}

A custom version of Lare3d~\citep{arberStaggeredGridLagrangian2001} has been
developed where a new module for anisotropic viscosity has been included. The
version including the new module can be found at
\url{https://github.com/jamiejquinn/Lare3d}, and has been archived
by~\revcite{\citet{keith_bennett_2020_4155546}}. The version of Lare3d used in the
production of the results presented here, including initial conditions,
boundary conditions, control parameters and the anisotropic viscosity module,
can be found in the repository of ~\revcite{\citet{keith_bennett_2020_4155625}}. Associated running scripts for generating, building and running simulations on a cluster is also provided~\citet{jamie_j_quinn_2022_6327300}. The data analysis and
instructions for reproducing all results found in this report may be also found
at \url{https://github.com/JamieJQuinn/coronal-fluting-instability-analysis}
and has been
archived~\citep{quinnJamieJQuinnCoronalflutinginstabilityanalysis2021}.

All simulations were performed on a single, multi-core machine with $40$ cores
provided by Intel Xeon Gold 6138 Skylake processor running at $2$ GHz and $192$
GB of RAM, although this amount of RAM is much higher than was required;
a conservative estimate of the memory used in the largest simulations is around
$64$ GB. Most simulations completed in under $2$ days.

\bsp	
\label{lastpage}
\end{document}